\documentclass[twocolumn,A4]{webofc}
\usepackage{graphicx} 
\usepackage[varg]{txfonts}   
\usepackage{hyperref}
\usepackage{url}
\hypersetup{colorlinks=true,citecolor=blue,urlcolor=blue,linkcolor=blue}
\usepackage[normalem]{ulem}
\usepackage{color}
\def\bfm{\bf\boldmath}

\def\be{\begin{equation}}
\def\ee{\end{equation}}
\def\bea{\begin{eqnarray}}
\def\eea{\end{eqnarray}}
\def\nn{\begin{nonumber}}

\begin{document}
%


\title{
\vspace*{-4ex}
\begin{flushright}
\hfill {\small LFTC-25-10/104}
\end{flushright}
\vspace{2ex}
In-medium mass shifts of $B_c^{(*)}, B_s^{(*)}$ and $D_s^{(*)}$ mesons}


\author{\firstname{Kazuo} \lastname{Tsushima}\inst{1}\fnsep
\thanks{\email{kazuo.tsushima@gmail.com}}
\thanks{\email{kazuo.tsushima@cruzeirodosul.edu.br}}
\and
\firstname{Samuel} \lastname{Beres}\inst{1}\fnsep
\thanks{\email{samuel.beres@hotmail.com}}
\and
\firstname{Guilherme} \lastname{Zeminiani}\inst{1}\fnsep
\thanks{\email{guilherme.zeminiani@gmail.com}}
}

\institute{Laboratório de Física Teórica e Computacional (LFTC), Programa de
P\'{o}sgradua\c{c}\~{a}o em Astrof\'{i}sica e F\'{i}sica Computacional,
Universidade Cidade de S\~{a}o Paulo (UNICID), 01506-000 S\~{a}o Paulo, SP, Brazil
}

\abstract{
We present our predictions for the Lorentz scalar mass shifts of
two-flavored heavy mesons, $B_c^{(*)}, B_s^{(*)}$ and $D_s^{(*)}$ in symmetric
nuclear matter. The in-medium mass shifts are estimated by evaluating
the lowest order one-loop self-energies of the mesons based on a flavor-SU(5)
effective Lagrangian approach.
In-medium properties necessary for the estimates are calculated by
the quark-meson coupling (QMC) model.
The enhanced self-energies of the mesons in symmetric nuclear matter relative to those
in free space, yield the negative mass shifts of these mesons.
}
\maketitle

\section{Introduction}
\label{intro}

Studies of hadron interactions with nuclear medium,
particularly for the mesons composed of two-flavored heavy quarks,
can shed light on the roles of gluons in quantum chromodynamics (QCD).
Since such mesons do not share light quarks with the nucleons (nuclear medium),
the interactions with the nuclear medium are expected primarily by gluons at the lowest order.
Here, our focus is on the mesons $B_c^{(*)}, B_s^{(*)}$ and $D_s^{(*)}$,
and their mass shifts in symmetric nuclear matter based on Ref.~\cite{Zeminiani:2023gqc}.
(See, e.g.,
Refs.~\cite{Tsushima:1998ru,Krein:2010vp,Krein:2017usp,Cobos-Martinez:2020ynh,Zeminiani:2024dyo,
Cobos-Martinez:2025iqg}
for the studies made on the mass shifts of quarkonia and heavy-meson-nucleus bound states.)

\begin{figure}[htb]
\centering
\includegraphics[scale=0.224]{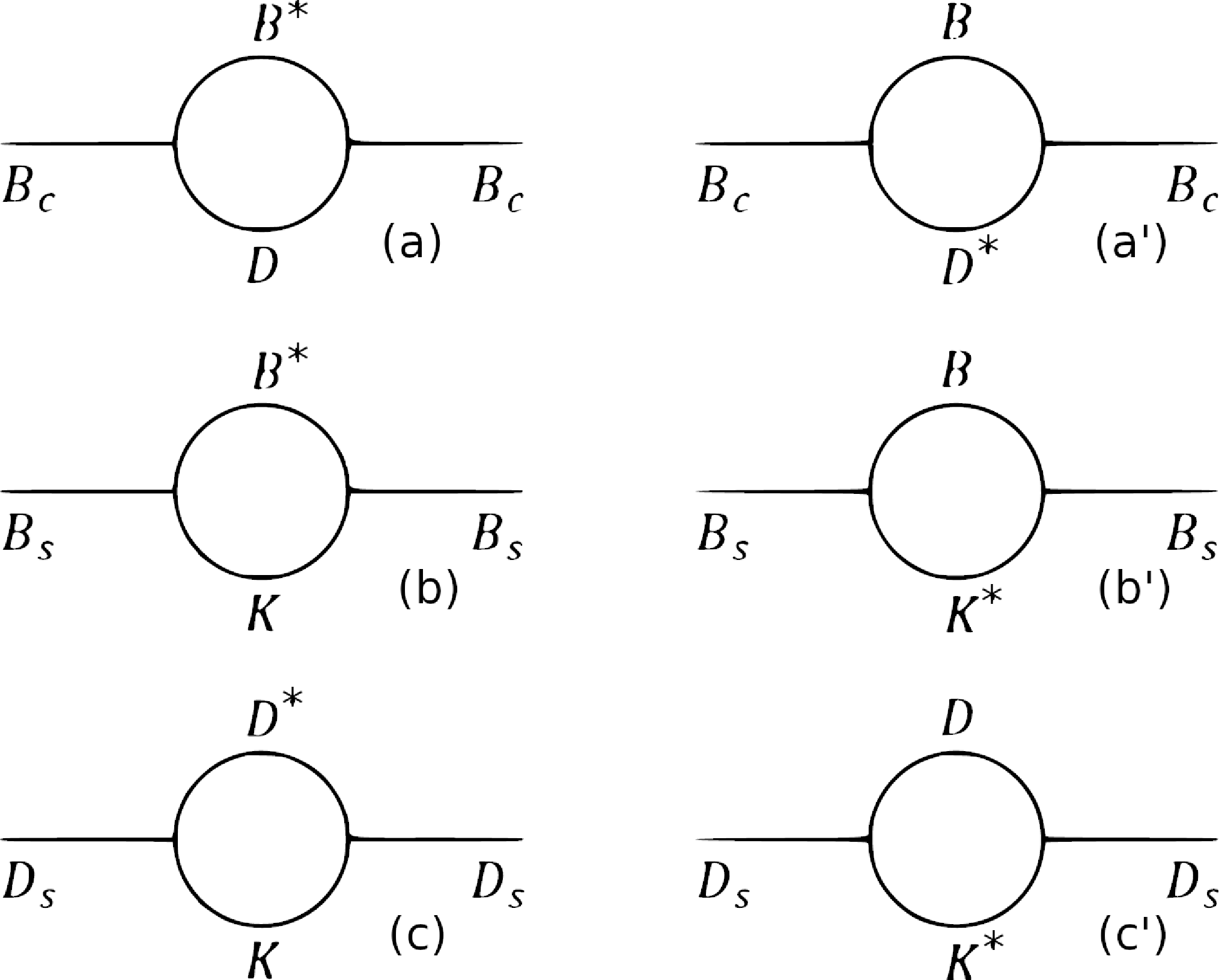}
\hspace{1.0cm}
\includegraphics[scale=0.31]{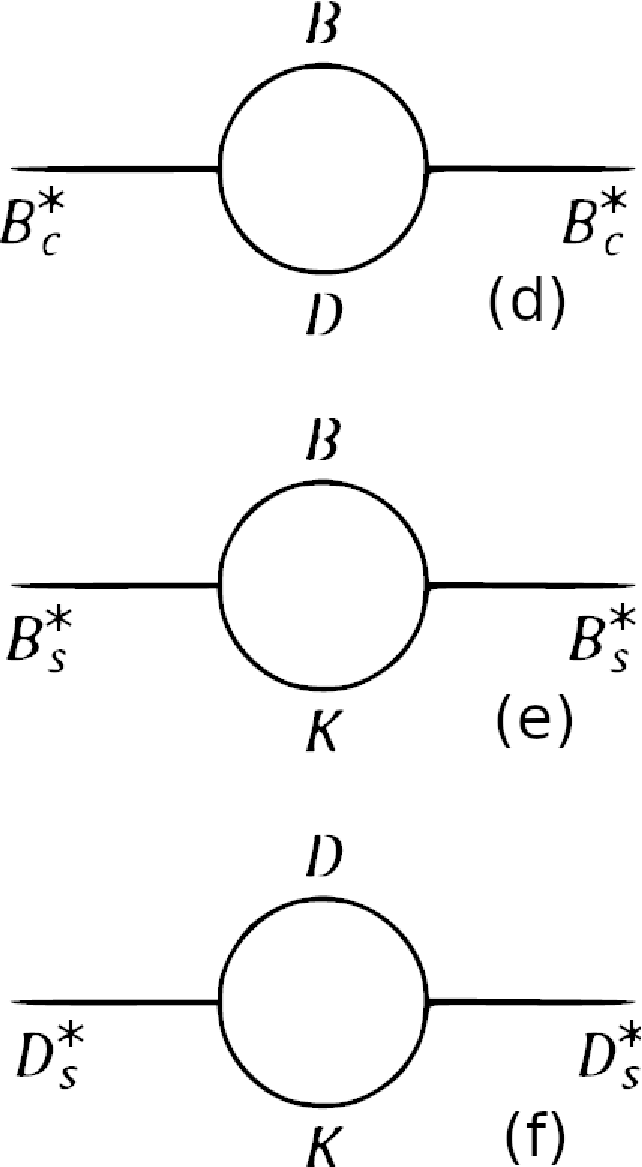}
\caption{Meson self-energy diagrams for mesons $B_c$ [(a) and (a')],
$B_s$ [(b) and (b')], $D_s$ [(c) and (c')],
$B_{c}^{*}$ [(d)], $B_{s}^{*}$ [(e)],
and $D_{s}^{*}$ [(f)], included in the present study.
\label{psvmeson}
\vspace{-3ex}
}
\end{figure}

In Fig.~\ref{psvmeson}, we show the self-energy graphs
included for evaluating the mass-shifts of $B_c^{(*)}, B_s^{(*)}$
and $D_s^{(*)}$ mesons.
(Note that, for the self-energies of the vector mesons $B_c^*, B_s^*$ and $D_s^*$, we
include the ''lowest order graphs'' excluding the two-vector-meson excited
intermediate states, based on the discussions given in Ref.~\cite{Zeminiani:2020aho}.)

To evaluate the in-medium self-energy graphs depicted in Fig.~\ref{psvmeson},
we use the in-medium masses of the intermediate state
excited mesons, $B^{(*)}, D^{(*)}$ and $K^{(*)}$,
calculated by the quark-meson coupling (QMC) model.
Thus, we will briefly explain next the QMC model~\cite{Guichon:1987jp,Saito:2005rv}.

\section{Quark-meson coupling (QMC) model}
\label{secqmc}

The Lorentz-scalar and Lorentz-vector mean field potentials in symmetric nuclear matter
for the $B^{(*)}, D^{(*)}$ and $K^{(*)}$ mesons enter to evaluate the
self-energy graphs shown in Fig.~\ref{psvmeson}, we use those calculated
by the QMC model~\cite{Guichon:1987jp,Saito:2005rv}.
In the QMC model, the binding of nucleons in nuclei (nuclear medium) is achieved by
the self-consistent exchange of the scalar-isoscalar-$\sigma$, vector-isoscalar-$\omega$ and
vector-isovector-$\rho$ meson fields directly coupled to the
light quarks $u$ and $d$, that are confined and relativistically moving inside the nucleon
(described by the MIT bag model).
The meson mean fields are generated by those confined light quarks in the nucleons,
and the (light-quark)-(meson field) coupling constants in any hadrons should be the same
as those in nucleon, after they are calibrated once properly.
Then, one can naturally expect that the hadrons with light quarks,
when they are immersed in nuclear medium, are expected to change their properties
in the QMC model.

In the rest frame of nuclear matter, the Dirac equations for the
quarks and antiquarks (no Coulomb force) in the QMC model,
assuming SU(2) symmetry for the light quarks ($q=u,d$ and $m_q \equiv m_u = m_d$)
are given by~\cite{Tsushima:1997df},
\begin{eqnarray}
&&\hspace{-8ex}\left[i\gamma \cdot \partial_{x} - \left(m_{q} - V^{q}_{\sigma}\right)
\mp \gamma^{0} \left(V^{q}_{\omega} + \frac{1}{2}V^{q}_{\rho}\right)\right]
\begin{pmatrix}
        \psi_{u}\left(x\right)\\
        \psi_{\overline{u}}\left(x\right)
       \end{pmatrix} = 0,
\label{Dequ}\\
&&\hspace{-8ex}\left[i\gamma \cdot \partial_{x} - \left(m_{q} - V^{q}_{\sigma}\right)
\mp \gamma^{0} \left(V^{q}_{\omega} - \frac{1}{2}V^{q}_{\rho}\right)\right]
\begin{pmatrix}
        \psi_{d}\left(x\right)\\
        \psi_{\overline{d}}\left(x\right)
       \end{pmatrix} = 0,
\label{Deqd}\\
&&\hspace{-8ex}\left[i\gamma \cdot \partial_{x} - m_{Q}\right]\psi_{Q, \overline{Q}}\left(x\right) =
0,
\end{eqnarray}
where $Q = s,c$, or $b$
and the mean-field potentials for the light quark ($q$) in nuclear matter are defined by
$V^{q}_{\sigma} \equiv g^{q}_{\sigma}\sigma$,
$V^{q}_{\omega} \equiv g^{q}_{\omega}\omega = g^q_\omega\, \delta^{\mu,0} \omega^\mu$,
$V^{q}_{\rho} \equiv g^{q}_{\rho}b = g^q_\rho\, \delta^{i,3} \delta^{\mu,0} \rho^{i,\mu}$,
with $\sigma$, $\omega$ and $b$ being the meson mean fields, and
$g^{q}_{\sigma}$, $g^{q}_{\omega}$ and $g^{q}_{\rho}$ the corresponding
quark-meson coupling constants.
Hereafter, we consider symmetric nuclear matter in Hartree approximation, and
set $V^q_\rho = 0$ in Eqs.~(\ref{Dequ}) and~(\ref{Deqd}).

The eigenenergies for the quarks and antiquarks in a hadron
$h\, (=\hspace{-1ex}B,B^*, D, D^*, K, \rm or\, K^*)$ in units of
the in-medium bag radius of hadron $h$, $1/R^{*}_{h}$
\footnote{The in-medium quantity is indicated
by an asterisk '*', except for indicating the vector mesons,
$B^*, D^*, K^*, B_c^*, B_s^*$ and $D_s^*$.}
are given by (recall that $V^q_\rho=0$ in symmetric nuclear matter below):
\begin{eqnarray}
&&\begin{pmatrix}
        \epsilon_{u}\\
        \epsilon_{\overline{u}}
       \end{pmatrix} = \Omega^{*}_{q} \pm R^{*}_{h}
\left(V^{q}_{\omega} + \frac{1}{2}V^{q}_{\rho}\right),\\
&&\begin{pmatrix}
        \epsilon_{d}\\
        \epsilon_{\overline{d}}
       \end{pmatrix} = \Omega^{*}_{q} \pm R^{*}_{h}
\left(V^{q}_{\omega} - \frac{1}{2}V^{q}_{\rho}\right),\\
&&\epsilon_{s,c,b} = \epsilon_{\overline{s},\overline{c},\overline{b}} =
\Omega^*_{s,c,b}.
\end{eqnarray}

The in-medium mass of the hadron $h$, $m^*_h$, for a given nuclear density
is calculated by
\begin{equation}
\hspace{-2ex}
m^{*}_{h} = \sum_{j=q,\overline{q},Q,\overline{Q}}
\frac{n_{j}\Omega^{*}_{j}- Z_{h}}{R^{*}_{h}} + \frac{4}{3} \pi R^{*3}_{h}B_{p},
\hspace{2ex}
\left. \frac{d m^{*}_{h}}{d R_{h}}\right|_{R_{h} = R^{*}_{h}} = 0,
\end{equation}
with $\Omega^{*}_{q} = \Omega^{*}_{\overline{q}} = \left[x^{2}_{q}
+ \left(R^{*}_h m^{*}_{q}\right)^{2}\right]^{\frac{1}{2}}$,
where $m^{*}_{q} = m_{q} - V^q_\sigma$
and $\Omega^{*}_{Q} = \Omega^{*}_{\overline{Q}} = \left[x^{2}_{Q} + \left(R^{*}_{h}
m_{Q}\right)^{2}\right]^{\frac{1}{2}}$ when $h$ contains a light quark
(when $h$ contains no light quarks, $R_h^* \to R_h$),
and $x_{q,Q}$ are the lowest mode bag eigenfrequencies.
Note that, when $h$ contains a light quark, $x_Q$ is also modified slightly from
the free space value due to the in-medium modification of $R^*_h$,
and thus $\Omega_Q$ is also modified as $\Omega^*_Q$, from that in free space.
$B_{p}$ is the bag constant, and $n_{q,Q}$ ($n_{\overline{q},\overline{Q}}$) are the lowest
mode quark (antiquark) numbers for the quark flavors $q$ and $Q$ in the hadron $h$,
and the $Z_{h}$ parametrizes the sum of the center-of-mass
and gluon fluctuation effects, which is assumed to be independent
of density~\cite{Guichon:1995ue}.

The current quark mass values used are
$(m_q, m_s, m_c, m_b)$ = (5, 250, 1270, 4200) MeV.
(See Ref.~\cite{Tsushima:2020gun} for the other values used,
$(m_q, m_s, m_c, m_b)$ = (5, 93, 1270, 4180) MeV.)
Note that, in phenomenological quark models, the quark mass values
are not necessarily related with the values in quantum chromodynamics (QCD).
The free space nucleon bag radius is chosen to be
$R_{N}$ = 0.8 fm, and the light quark-meson coupling constants,
$g^{q}_{\sigma}$, $g^{q}_{\omega}$ and $g^{q}_{\rho}$, are determined by
the fit to the symmetric saturation energy (-15.7 MeV)
at the saturation density ($\rho_0 = 0.15$ fm$^{-3}$), and the bulk symmetry energy
(35 MeV), and the explicit parameter values obtained are
given in Refs.~~\cite{Guichon:1987jp,Saito:2005rv,Tsushima:2020gun}.

In Table~\ref{mesonmass} we summarize the free space meson mass values (input)
taken from Particle Data Group (PDG)~\cite{ParticleDataGroup:2022pth}
except for that of the $m_{B^*_c}$ which is not determined experimentally,
and for this, we use the average value ofo
$m_{B^*_c}$ from Table III in Ref.~\cite{Martin-Gonzalez:2022qwd}.
We also list the in-medium mass values for some mesons appearing
in Fig.~\ref{psvmeson}, at $\rho_0, 2\rho_0$ and $3\rho_0$ ($\rho_0=0.15$ fm$^{-3}$)
calculated by the QMC model. The entries with the bold face are the focus of
the present study.
\begin{table}[htb!]
\caption{\label{mesonmass}
Free space meson mass values (2nd column, input) from Particle Data Group
(PDG)~\cite{ParticleDataGroup:2022pth}, and in-medium masses for some mesons,
at densities $\rho_0$, $2\rho_0$ and $3\rho_0$ with $\rho_0$ = 0.15 fm$^{-3}$
calculated by the QMC model
with $(m_{u,d}$,$m_s$,$m_c$,$m_b)$ = (5,250,1270,4200)
MeV\cite{Tsushima:2020gun}.
(For $m_{B^*_c}$ we use the averaged value from Table III
in Ref.~\cite{Martin-Gonzalez:2022qwd}.)
The entries with bold face are those for the two-flavored heavy mesons,
that we focus on the present study. (All units are in MeV.)
}
\label{mesonmass}
\vspace{1ex}
\begin{tabular}{l|c|c|c|c}
\hline
\hline
           &$\rho_B=0$ &$\rho_B=\rho_0$ &$\rho_B=2\rho_0$
&$\rho_B=3\rho_0$\\
\hline
\hline
$m_K$       &493.7  &430.5  &393.6  &369.0  \\
$m_{K^*}$   &893.9  &831.9  &797.2  &775.0  \\
$m_D$       &1867.2 &1805.2 &1770.6 &1748.4 \\
$m_{D^*}$   &2008.6 &1946.9 &1912.9 &1891.2 \\
$m_B$       &5279.3 &5218.2 &5185.1 &5164.4 \\
$m_{B^*}$   &5324.7 &5263.7 &5230.7 &5210.2 \\
$\bfm m_{B_{c}}$   &6274.5  &  &  &  \\
$\bfm m_{B^{*}_{c}}$   &6333.0  &  &  &  \\
$\bfm m_{B^0_{s}}$   &5366.9  &  &  &  \\
$\bfm m_{B^{*}_{s}} (\equiv m_{B^{* 0}_{s}})$   &5415.4  &  &  &  \\
$\bfm m_{D_{s}}$ &1968.4  &  &  &  \\
$\bfm m_{D^{*}_{s}}$ &2112.2  &  &  &  \\
\hline
\hline
\end{tabular}
\vspace{-2ex}
\end{table}
%

In the QMC model the density dependent negative mass shifts of the
$B$, $B^*$, $D$, $D^*$, $K$ and $K^*$ mesons
are obtained~\cite{Tsushima:2020gun,Zeminiani:2020aho,Zeminiani:2023gqc}.
The model predicts the mass shifts ($\Delta m$ $\equiv$ $m^*-m$)
of these mesons in symmetric nuclear matter at $\rho_0 = 0.15$ fm$^{-3}$:
$(\Delta m_B, \Delta m_{B^*},
\Delta m_D, \Delta m_{D^*},
\Delta m_K, \Delta m_{K^*})$ =
(-61.13, -61.05, -61.97, -61.66, -63.20, -61.97) MeV.
We use the density dependent in-medium masses of these mesons
for estimating the $B_c$, $B^*_c$, $B_s$, $B^*_s$, $D_s$
and $D^*_s$ meson self-energies in symmetric nuclear matter, and since
the vector potentials cancel out in the intermediate excited state mesons,
we do not need them.
\begin{figure}[htb]
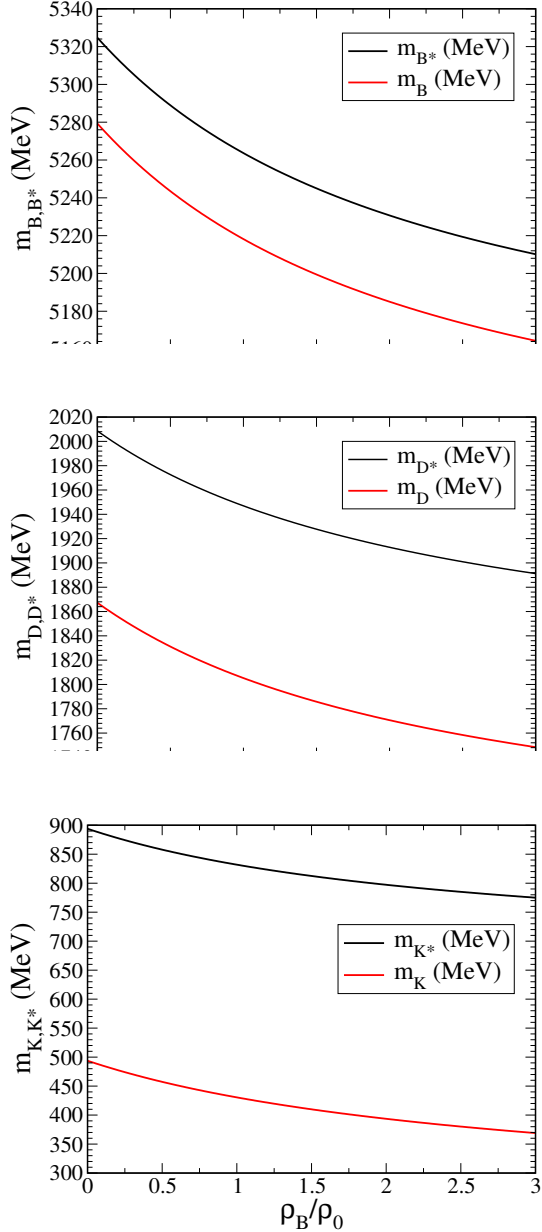
%
\centering
\includegraphics[width=7.0cm]{meson_BBs_mass.eps}
\includegraphics[width=7.0cm]{meson_DDs_mass.eps}
\includegraphics[width=7.0cm]{meson_KKs_mass.eps}
\caption{$B$ and $B^*$ (top), $D$ and $D^*$ (middle) and $K$ and $K^*$
(middle) meson Lorentz-scalar effective masses in symmetric nuclear matter
versus baryon density ($\rho_B/\rho_0$), calculated by the QMC model.
\vspace{-2ex}}%
\label{bksmass}
\end{figure}

\section{Effective Lagrangian approach for two-flavored heavy meson mass shift}
\label{secmshft}

In the following, we mainly focus on the $B_c$ and $B_c^*$ mesons.
The $B_c$ [$B^*_c$] in-medium mass shift arises from
the $B^*D + BD^*$ [$BD$] loop contribution to the self-energy,
relative to those in free space (see figures (a) and (a') [(d)] in Fig.~\ref{psvmeson}).

The self-energies are calculated based on a flavor-SU(5) symmetric effective Lagrangian
densities~\cite{Lin:2000ke}. (Hereafter we will simply call ''Lagrangian''.)
The free Lagrangian for pseudoscalar and vector mesons
(denoted respectively by $P$ and $V$ of $5 \times 5$
matrix representations to be shown later) is given by~\cite{Lin:2000ke},
\begin{equation}
{\cal L}_{0}=Tr \left(  \partial _{ \mu }P^{\dagger} \partial ^{ \mu }P \right)
-\frac{1}{2}Tr \left( F_{ \mu  \nu }^{\dagger}F^{ \mu  \nu } \right),
\end{equation}
with $F_{ \mu  \nu }= \partial _{ \mu }V_{ \nu }- \partial _{ \nu }V_{ \mu }$,
where $P$ and $V$ (suppressing the Lorentz indices for $V$) are, respectively,
the $5 \times 5$ pseudoscalar and vector meson matrices in SU(5) and given below
(for a clearer identification of the matrix elements,
see, e.g., Refs.~\cite{Lin:2000ke,Zeminiani:2020aho,Zeminiani:2023gqc}):
{\tiny
\begin{eqnarray}
&&\hspace{-28ex} P = \frac{1}{\sqrt{2}} \begin{pmatrix} 
\frac{\pi^{0}}{\sqrt{2}} + \frac{\eta}{\sqrt{6}} + \frac{\eta_{c}}{\sqrt{12}}
+ \frac{\eta_{b}}{\sqrt{20}}  &  \pi^{+}  &  K^{+}  &  \overline{D}^{0}  &  B^{+}\\
\pi^{-}  &  \frac{-\pi^{0}}{\sqrt{2}} + \frac{\eta}{\sqrt{6}} + \frac{\eta_{c}}{\sqrt{12}}
+ \frac{\eta_{b}}{\sqrt{20}}  &  K^{0}  &  D^{-}  &  B^{0}\\
K^{-}  &  \overline{K}^{0}  &  \frac{-2\eta}{\sqrt{6}} + \frac{\eta_{c}}{\sqrt{12}}
+ \frac{\eta_{b}}{\sqrt{20}}  &  D_{s}^{-}  &  B_{s}^{0}\\
D^{0}  &  D^{+}  &  D_{s}^{+}  &  \frac{-3\eta_{c}}{\sqrt{12}}
+ \frac{\eta_{b}}{\sqrt{20}}  &  B_{c}^{+}\\
B^{-}  &  \overline{B^{0}}  &  \overline{B_{s}^{0}}  &  B_{c}^{-}
&  \frac{-2\eta_{b}}{\sqrt{5}}\\
\end{pmatrix}, \nonumber \label{p} \\
\nonumber\\
\nonumber\\
&&\hspace{-28ex} V = \frac{1}{\sqrt{2}} \begin{pmatrix}
\frac{\rho^{0}}{\sqrt{2}} + \frac{\omega}{\sqrt{6}} + \frac{J/\Psi}{\sqrt{12}}
+ \frac{\Upsilon}{\sqrt{20}}  &  \rho^{+}  &  K^{*+}  &  \overline{D}^{*0}  &  B^{*+}\\
\rho^{-}  &  \frac{-\rho^{0}}{\sqrt{2}} + \frac{\omega}{\sqrt{6}} + \frac{J/\Psi}{\sqrt{12}}
+ \frac{\Upsilon}{\sqrt{20}}  &  K^{*0}  &  D^{*-}  &  B^{*0}\\
K^{*-}  &  \overline{K}^{*0}  &  \frac{-2\omega}{\sqrt{6}} + \frac{J/\Psi}{\sqrt{12}}
+ \frac{\Upsilon}{\sqrt{20}}  &  D_{s}^{*-}  &  B_{s}^{*0}\\
D^{*0}  &  D^{*+}  &  D_{s}^{*+}  &  \frac{-3J/\Psi}{\sqrt{12}} + \frac{\Upsilon}{\sqrt{20}}  &
B_{c}^{*+}\\
B^{*-}  &  \overline{B^{*0}}  &  \overline{B_{s}^{*0}}  &  B_{c}^{*-}  &
\frac{-2\Upsilon}{\sqrt{5}} \\
\end{pmatrix}. \nonumber \label{v}
\end{eqnarray}
}

By the following minimal substitutions,
\begin{eqnarray}
&&\partial _{ \mu }P \rightarrow  \partial _{ \mu }P-\frac{ig}{2} \left[ V_{ \mu }\text{, P}
\right],\\
&&F_{ \mu  \nu } \rightarrow  \partial _{ \mu }V_{ \nu }- \partial _{ \nu }V_{ \mu }
 -\frac{ig}{2} \left[ V_{ \mu },~V_{ \nu } \right],
\end{eqnarray}
one obtains the effective interaction Lagrangians~\cite{Lin:2000ke}.
For the $B_c B^*D$, $B_c BD^*$ and $B^*_c BD$ interactions
the relevant Lagrangians are~\cite{Zeminiani:2023gqc,Lodhi:2007zz}:
\begin{eqnarray}
\hspace{-6ex}
    \mathcal{L}_{B_{c}B^{*}D} &=& ig_{B_{c}B^{*}D}
    [(\partial_{\mu}B^{-}_{c}){D}
    - B^{-}_{c}(\partial_{\mu}{D})] B^{* \mu} + h.c.,
    \nonumber \\
\hspace{-6ex}
    \mathcal{L}_{B_{c}BD^{*}} &=& ig_{B_{c}BD^{*}}
    [(\partial_{\mu}B^{+}_{c})\overline{B}
    - B_c^{+}(\partial_{\mu}\overline{B})] \overline{D^{*}}^\mu + h.c.,
   \nonumber \\
\hspace{-6ex}
   \mathcal{L}_{B^{*}_{c}BD} &=& -ig_{B^{*}_{c}BD}
     B^{*+{\mu}}_{c} [\overline{B}(\partial_{\mu} \overline{D}) -
(\partial_{\mu} \overline{B}) \overline{D}] + h.c.,
\label{lagBc}
\end{eqnarray}
where the following conventions are used with the superscript $T$
standing for the ''transposition'' operation:
\begin{align*}
B&=\begin{pmatrix}
B^{+}\\B^{0}
\end{pmatrix},
&\hspace{-2ex}\overline{B}=\begin{pmatrix}
B^{-}\\ \overline{B}^{0}
\end{pmatrix}^T,
&\hspace{1ex}B^{*} =\begin{pmatrix}
B^{*+}\\B^{*0}
\end{pmatrix},
&\hspace{-2ex}\overline{B^{*}}&=\begin{pmatrix}
B^{*-}\\ \overline{B}^{*0}
\end{pmatrix}^T,
\end{align*}
\begin{align*}
\overline{D}&=\begin{pmatrix}
\overline{D}^{0}\\D^{-}
\end{pmatrix},
&\hspace{-2ex}D=\begin{pmatrix}
D^{0}\\D^{+}
\end{pmatrix}^T,
&\hspace{1ex}\overline{D}^{*} =\begin{pmatrix}
\overline{D}^{*0}\\D^{*-}
\end{pmatrix},
&\hspace{-2ex}D^{*}&=\begin{pmatrix}
D^{*0}\\ D^{*+}
\end{pmatrix}^T.
\end{align*}

The universal coupling constant $g$ appearing in the flavor
SU(5) Lagrangian has the following relation, and also the value is fixed
as~\cite{Zeminiani:2023gqc},
$g_{\Upsilon BB} = \frac{5g}{4\sqrt{10}} \approx 13.2$,
using the $\Upsilon$ decay data $\Gamma(\Upsilon \to e^+ e^-)$
and the vector meson dominance (VMD) model~\cite{Lin:2000ke,Zeminiani:2020aho}.
The coupling constant values appearing in Eq.~(\ref{lagBc}) are fixed by
the flavor SU(5) symmetry:
\bea
g_{B_c B^* D} &=& \frac{2}{\sqrt{5}}g_{\Upsilon BB},
\\
g_{B_c B^* D} &=& g_{B_c B D^*} = g_{B^*_c B D} = \frac{g}{2\sqrt{2}}
\approx 11.9.
\eea

The in-medium $B_c$ mass shift, $\Delta m_{B_c}$ (Lorentz-scalar),
is computed by the difference of the in-medium $m^*_{B_c}$ and the free space
$m_{B_c}$ masses as,
\begin{equation}
\Delta m_{B_c} = m^*_{B_c} - m_{B_c},
\end{equation}
where, the free space mass $m_{B_c}$ (input) is used to determine the bare mass $m^0_{B_c}$
with the $B_c$ self-energy $\Sigma_{B_c}$ by
\begin{equation}
m^2_{B_c} = \left( m^0_{B_c} \right)^2 -
{\displaystyle |}\Sigma_{B_c}(k^2 = m^2_{B_c}){\displaystyle |}.
\label{m0}
\end{equation}
Note that, the total self-energy $\Sigma_{B_c}$
is calculated by the ($B^*D$ + $BD^*$) meson loop contribution in free space,
ignoring any possible $B_c$ meson width as well as those of the other mesons
(or imaginary part) in the self-energy.
The in-medium $B_c$ mass $m^{* 2}_{B_c}$
is similarly calculated using the same bare mass
value $m^0_{B_c}$ determined in free space by reproducing the observed mass,
using the density dependent in-medium masses of the
($B,B^*,D,D^*$) mesons ($m^*_{B},m^*_{B^*},m^*_{D},m^*_{D^*}$), namely,
\bea
\hspace{-2ex}
m^2_{B_c}
&=& \left[ m^0_{B_c}(B^*D+BD^*) \right]^2
\nonumber\\
&&- \left|\Sigma_{B_c}(B^*D) + \Sigma_{B_c}(BD^*) \right|(k^2 = m^2_{B_c}),
\label{m02}\\
\hspace{-2ex}
m^{* 2}_{B_c}
&=& \left[ m^0_{B_c}(B^*D+BD^*) \right]^2
\nonumber\\
&&- \left|\Sigma^*_{B_c}(B^*D) + \Sigma^*_{B_c}(BD^*) \right|(k^{* 2} = m^{* 2}_{B_c}).
\label{m03}
\eea

We note that, when the self-energy graphs contain different contributions as
the present case as $\Sigma_{B_c} ({\rm total}) = \Sigma(B^*D)$ + $\Sigma(BD^*)$ in free space,
$m^0$ depends on both $\Sigma(B^*D)$ and $\Sigma(BD^*)$ in reproducing the physical mass
$m_{B_c}$. Thus, one must be careful when discussing the $B_c$
in-medium mass and mass shift of each loop contribution
$\Sigma(B^*D)$ and $\Sigma(BD^*)$,
since $m^0(B^*D+BD^*) \ne m^0(B^*D) \ne m^0(BD^*)$,
and $m^0(B^*D + BD^*) \ne m^0(B^*D) + m^0(BD^*)$.
The dominant loop contribution should be known by the decomposition
of the total contribution in the self-energy,
$\Sigma^{(*)}_{B_c} (B^*D + BD^*) = \Sigma^{(*)}_{B_c}(B^*D) + \Sigma^{(*)}_{B_c}(BD^*)$.

We focus on the $B^*D$ loop contribution now, without that of the $BD^*$ loop.
The $B_c$ self-energy is calculated by
\begin{equation}
    \Sigma^{B^{*}D}_{B_{c}}(m^{*}_{B_{c}}) = \frac {-4g^{2}_{B_{c}B^{*}D}}{\pi^{2}}
    \int d|\textbf{k}|\,
    |\textbf{k}|^{2} I_{B_c}^{B^{*}D}(|\textbf{k}|)
    F_{B_c B^* D} (\textbf{k}^2),
\label{SigBc}
\end{equation}
where $I_{B_c}^{B^{*}D} (|\textbf{k}|)$ is expressed as below,
after the Cauchy integral with respect to $k^0$ complex plane by shifting $k^0$ variable
for the vector potential:
\begin{eqnarray}
&&\hspace{-10ex}I_{B_c}^{B^{*}D} (|\textbf{k}|) =
\nonumber\\
&&\hspace{-8ex}\left. \frac{m^{*2}_{B_c} \left(-1 + k^2_0 / m^{*2}_{B^*} \right)}
{(k_0 - \omega^*_{B^*}) (k_0 - m^{*}_{B_c} + \omega^*_D)
(k_0 - m^{*}_{B_c} -\omega^*_D)}
\right|_{k_{0} = -\omega^*_{B^*} }
\nonumber \\
&& \left. \hspace{-8ex} + \frac{m^{*2}_{B_c} \left( -1 + k^2_0 / m^{*2}_{B^*} \right) }
{(k_0 + \omega^*_{B^*}) (k_0 - \omega^*_{B^*})
(k_0 - m^{*}_{B_c} -\omega^*_D)}
\right|_{k_{0} = m^{*}_{B_c} - \omega^*_D}.
\label{IBsD}
\end{eqnarray}
In Eq.~(\ref{SigBc}), $F_{B_c B^* D}$ is the product of vertex form factors
to regularize the divergence in the loop integral,
$F_{B_c B^* D} (\textbf{k}^2)$ $\equiv$ $u_{B_cB^*}(\textbf{k}^2) \times u_{B_c D}(\textbf{k}^2)$,
and $u_{B_cB^*}(\textbf{k}^2)$ and $u_{B_cD}(\textbf{k}^2)$ are respectively given by
$u_{B_c B^*}(\textbf{k}^2)$ = $\left( \frac{\Lambda^2_{B^*} + m^2_{B_c}}
{\Lambda^2_{B^*} + 4\omega^2_{B^*} (\textbf{k}^2)} \right)^2$ and
$u_{B_c D}(\textbf{k}^2)$ = $\left( \frac{\Lambda^2_{D} + m^2_{B_c}}
{\Lambda^2_{D} + 4\omega^2_{D} (\textbf{k}^2)} \right)^2$
with $\Lambda_{B^*}$ and $\Lambda_{D}$ being the corresponding cutoff masses
associated with the $B^*$ and $D$ mesons, respectively.
We use the common value $\Lambda \equiv \Lambda_{B^*} = \Lambda_{D}$.
A similar calculation can be performed for the $BD^*$ loop contribution
by replacing $(B^*,D) \to (B,D^*)$ in Eqs.~(\ref{SigBc}), and (\ref{IBsD})
(including the form factors).

Furthermore, the $B_c^*$ meson self-energy can be calculated straightforwardly.
(See Ref.~\cite{Zeminiani:2023gqc} for details.)

The choice of cutoff mass value has nonegligible impacts on
the calculated self-energy results.
In the present study, we use the common cutoff mass value
$\Lambda \equiv \Lambda_{B,B^*,D,D^*,K,K^*}$
for all the form factors appearing in the self-energy integrals, and
vary the $\Lambda$ value. The $\Lambda$ value may be associated
with the energies to probe the internal structure (finite sizes) of the mesons.
In the previous study~\cite{Zeminiani:2020aho} it was observed that when the value
of the cutoff mass becomes close to the masses of the mesons appearing in
the self-energies, such large cutoff mass value and those values larger than that
value did not make sense to serve as ''proper'' form factor.
This is because the Compton wavelengths of the corresponding
cutoff mass values reach near and/or smaller than those of the meson sizes,
and they do not represent the effect of the ''meson's finite size'' properly.
Therefore, we need to constrain the $\Lambda$ value in such
a way that the form factors reflect properly the finite sizes of the mesons.
Based on the heavy quark and heavy meson symmetry in QCD, it may be justified that
we use the same range of values for $\Lambda$ as it was practiced for
the cases of quarkonia~\cite{Zeminiani:2020aho}.
Namely, we use the values, $\Lambda$ = 2000, 3000, 4000, 5000 and 6000 MeV.

In Table~\ref{Bcm0} we give the cutoff mass ($\Lambda$) value dependence of the
bare mass $m^0$ in Eq.~(\ref{m0}) for different cases of the
self-energy calculations of $B_c$ meson, namely, only the $B^*D$ loop, only the $BD^*$ loop, and
$B^*D + BD^*$ loops.
\vspace{-2ex}
\begin{table}[htb]
\caption{
Cutoff $\Lambda$ value dependence of bare mass $m^0$ in Eq.~(\ref{m0})
for the cases of only the $B^*D$ loop, only the $BD^*$ loop, and $B^{*}D + BD^{*}$ loops.
(All numbers are in MeV).
\label{Bcm0}
}
\begin{center}   
\begin{tabular}{c|c|c|c}
\hline
\hline
$\Lambda$ &$m^0(B^{*}D)$ &$m^0(BD^{*})$ & $m^0(B^{*}D + BD^{*})$\\
\hline
\hline
2000		& 7906.1 &  9925.4 & 11029.6 \\
3000      	& 8032.6 & 10313.9 & 11468.7 \\
4000     	& 8249.0 & 10913.0 & 12156.0 \\
5000      	& 8561.5 & 11732.3 & 13098.8 \\
6000      	& 8968.6 & 12766.6 & 14284.6 \\
\hline
\hline
\end{tabular}
\end{center}
\end{table}
\vspace{-4ex}
One can notice from the results in Table~\ref{Bcm0} as follows.
First, the differences in the $m^0$ values for the different self-energy graphs
with the same cutoff $\Lambda$ value (each low) as already mentioned.
Second, the bare mass $m^0$ value increases
as the cutoff mass $\Lambda$ value increases (each column),
for all the individual cases of the self-energy contributions, $B^*D, BD^*$ and
$B^*D+BD^*$ loops.

Similarly, for the other two-flavored heavy mesons
$B_s, B_s^*, D_s$ and $D_s^*$,
with the same five $\Lambda$ values and the corresponding form factors,
we can estimate their in-medium mass shifts using the following
interaction Lagrangians~\cite{Zeminiani:2023gqc}:
\begin{eqnarray}
\hspace{-6ex}
    \mathcal{L}_{B_sB^{*}K} &=& ig_{B_sB^{*}K}[(\partial_{\mu}\overline{B^0_s})\overline{K}
     - \overline{B^0_s} (\partial_{\mu}\overline{K})]{B}^{* \mu} + h.c.,
     \nonumber \\
\hspace{-6ex}
     \mathcal{L}_{B_sBK^*} &=& ig_{B_sBK^*}[(\partial_{\mu}{B^0_s})\overline{B}
     - {B^0_s}(\partial_{\mu}\overline{B})]{K}^{* \mu} + h.c.,
     \nonumber \\
\hspace{-6ex}
     \mathcal{L}_{B^{*}_{s}BK} &=& -ig_{B^{*}_{s}BK}
     {B_s^*}^{0 \mu} [\overline{B} (\partial_\mu K)
     - (\partial_\mu \overline{B}) K]  + h.c.,
     \nonumber \\
\hspace{-6ex}
     \mathcal{L}_{D_sD^{*}K} &=& ig_{D_sD^{*}K}[(\partial_{\mu}D_{s}^+)\overline{K}
    - D_s^+ (\partial_\mu \overline{K})]\overline{D^*}^\mu + h.c.,
    \nonumber \\
\hspace{-6ex}
    \mathcal{L}_{D_sDK^*} &=& ig_{D_sDK^*}[(\partial_{\mu}D^-_{s})D
    - D_{s}^- (\partial_{\mu}D)  ]{K}^{* {\mu}} + h.c.,
    \nonumber \\
\hspace{-6ex}
    \mathcal{L}_{D^{*}_{s}DK} &=& -ig_{D^{*}_{s}DK}
     {D_s^{*-}}^\mu [D(\partial_{\mu}K)
     - (\partial_{\mu}D) K]  + h.c.
\end{eqnarray}
The coupling constant values associated with each vertex in the above, are obtained from
the universal SU(5) coupling $g$ value as,
%
\bea
\hspace{-8ex} g_{B_{s} KB^*} =&\hspace{-4ex}& g_{B_{s} BK^*} = g_{B^{*}_{s} BK} =
g_{D_{s} KD^*} = g_{D_{s} DK^*} = g_{D^{*}_{s} DK}
\nonumber \\
=&\hspace{-3ex}& \frac{g}{2\sqrt{2}} \approx 11.9.
\eea

\section{Mass shift results}

\begin{figure}[htb]
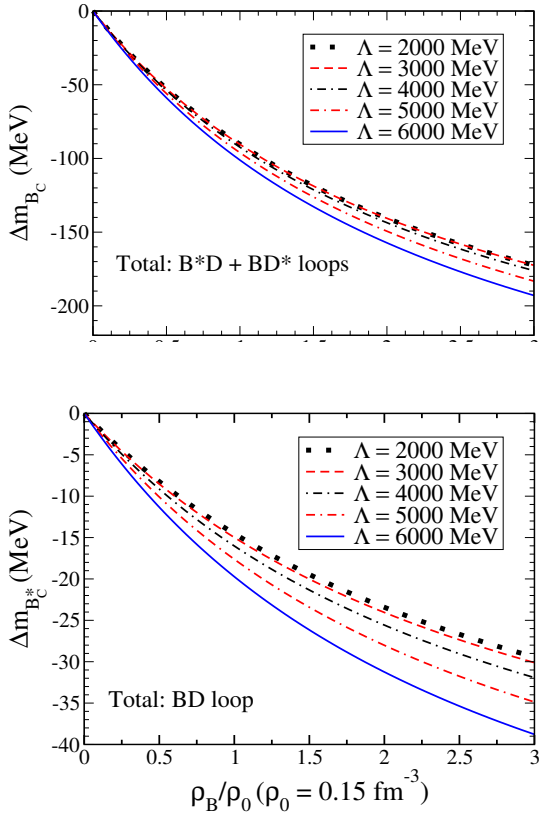

\centering
\includegraphics[width=7.0cm]{Bc_totalpot.eps}
\includegraphics[width=7.0cm]{Bcspot.eps}
\caption{The total ($B^*D$ + $BD^*$) loop contribution for the
in-medium $B_c$ mass shift (upper panel) and that of the $BD$ loop for $B_c^*$ (lower panel),
versus baryon density ($\rho_B/\rho_0$) for five different
cutoff mass values $\Lambda$.
\vspace{-2ex}}
\label{totbc}
\end{figure}
\begin{figure}[htb]
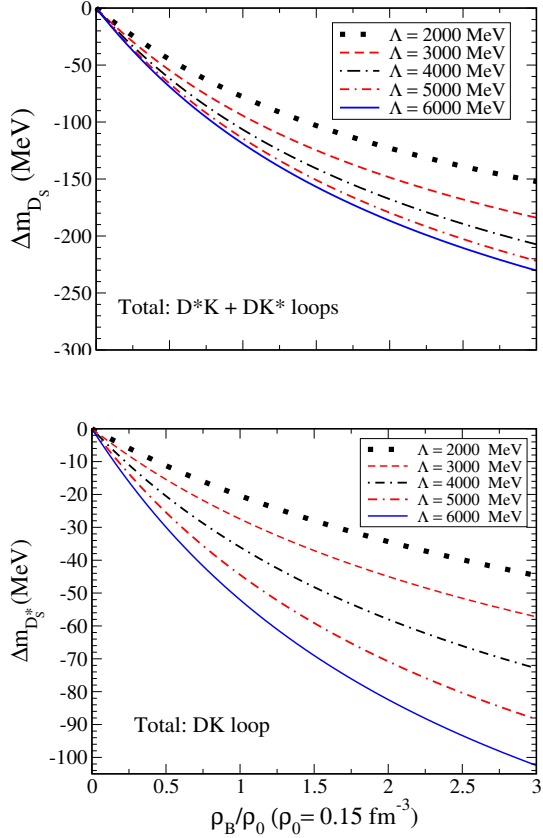

\centering
\includegraphics[width=7.0cm]{Dcs_totalpot.eps}
\includegraphics[width=7.0cm]{Dcsspot.eps}
\caption{The total ($D^*K$ + $DK^*$) loop contribution for the
in-medium $D_s$ mass shift (upper panel)
and that of the $DK$ loop for the $D_s^*$ (lower panel),
versus baryon density for five different
cutoff mass values~$\Lambda$.
\vspace{-2ex}}
\label{totds}
\end{figure}

We are now in a position to present our estimates (predictions) of
the in-medium mass shifts for the two-flavored heavy mesons,
($B_c, B_c^*$), ($B_s, B_s^*$), and ($D_s, D_s^*$)
mesons~\cite{Zeminiani:2023gqc} respectively in Figs.~\ref{totbc},~\ref{totbs0} and~\ref{totds}.

\begin{figure}[htb]
\centering
\includegraphics[width=7.0cm]{Bs0s_totalpot.eps}
\includegraphics[width=7.0cm]{Bsspotential.eps}
\caption{The total ($B^*K$+$BK^*$) loop contribution for the
in-medium $B_s^0$ mass shit (upper panel) and that of the $BK$ loop for the
$B_s^* \equiv B_s^{0 *}$ (lower panel),
versus baryon density ($\rho_B/\rho_0$) for five different
cutoff mass values $\Lambda$.
\vspace{-2ex}}
\label{totbs0}
\end{figure}

Among the all results given in Figs.~\ref{totbc},~\ref{totbs0} and~\ref{totds},
the cutoff mass value dependence of $\Delta m_{B_s^*}$
(the lower panel of Fig.~\ref{totbs0}) shows ''anomalous'' behavior.
Namely, $\Lambda = 2000$ MeV (thick dotted black line) gives
the most negative mass shift, despite all the other cases give the smallest
negative mass shift for each case with this value.
Furthermore, the mass-shift tendency (order) with the cutoff mass values
are also different from the ''typical one'', e.g., that shown in the lower panel of
Fig.~\ref{totds} -- as the cutoff mass value becomes larger, the more negative mass shift
is obtained.
This may be attributed that the mass difference in the intermediate state excited mesons
of $B$ and $K$, both in free space as well as in medium.
That is, the mass difference of them is the largest (see Table~\ref{mesonmass}),
and the role of $4\omega_K{\textbf{k}^2)}$ relative to $m_{B_c^*}^2$ in the form factor
$u_{B_c^*K}$ gives larger effects (in particular $\Lambda = 2000$ MeV)
than that of $4\omega_K{\textbf{k}^2)}$ relative to $m_{D_s^*}^2$ in
the form factor $u_{D_s^* K}$. The former tends to give larger
value than that of the latter, and thus larger self-energy enhancement
to yield larger negative mass shift.
But this analysis may not be sufficiently quantitative.

In addition to the density dependence of the mass shifts
shown in Figs.~\ref{totbc},~\ref{totbs0} and~\ref{totds},
we summarize below the mass shift ranges obtained at normal nuclear matter density
$\rho_0=0.15$ fm$^{-3}$ with the five different cutoff mass values in the form factors:
\\
(1) for $B_c$, $\Delta m_{B_c}$ = [-101.1, -90.4] MeV,
\\
(2) for $B^*_c$, $\Delta m_{B^*_c}$ = [-19.7, -14.5] MeV,
\\
(3) for $B_s$, $\Delta m_{B_s}$ = [-178.8, -133.0] MeV,
\\
(4) for $B^*_s$, $\Delta m_{B^*_s}$ = [-20.5, -16.0] MeV,
\\
(5) for $D_s$, $\Delta m_{D_s}$ = [-119.0, -78.0] MeV,
\\
(6) for $D^*_s$, $\Delta m_{D^*_s}$ = [-20.2, -52.1] MeV.

\section{Summary and Conclusions}
\label{seconcl}

We have presented our predictions for the in-medium mass shifts of two-flavored heavy mesons,
$B_c, B_c^*, B_s, B_s^*, D_s$ and $D_s^*$ in symmetric nuclear matter.
The estimates are made by evaluating the lowest order one-loop self-energy graphs
at the hadronic level, using a flavor-SU(5) effective Lagrangian approach,
with the in-medium inputs calculated by the quark-meson coupling (QMC) model.
We have not included any possible meson widths nor imaginary parts of the potentials.
Our results show the negative mass shifts for all the mesons studied,
$B_c, B_c^*, B_s, B_s^*, D_s$ and $D_s^*$.
In the evaluation, we have used some specific form factors to regularize the divergent
loop integrals with five values of the cutoff mass.
This part may be improved or has merit to study further in the future.

The next, interesting and important step, is to study possibilities of meson-nucleus bound
states for these mesons including the Coulomb potentials and/or the in-medium widths.
Concerning this, we would like to comment that, some initial results
for the $B_c$-$^{12}C$ system single-particle energies are already available in
Ref.~\cite{Zeminiani:2024dyo}.

\vspace{2ex}
\noindent
{\it Acknowledgments}:\hspace{1ex}
G.N.Z and K.T. thank the OMEG (Origin of Matter and Evolution of Galaxies)
Institute at Soongsil University (Seoul, South Korea) for the supports
and the collaboration works.
G.N.Z~and S.L.P.G.B.~were supported by the Coordena\c{c}\~ao de Aperfei\c{c}oamento
de Pessoal de N\'ivel Superior (CAPES), Brazil.
K.T.~was supported by Conselho Nacional de Desenvolvimento
Cient\'{i}fico e Tecnol\'ogico (CNPq), Brazil, Processes No.~304199/2022-2,
and
Funda\c{c}\~{a}o de Amparo à Pesquisa do Estado de S\~{a}o Paulo
(FAPESP), Brazil, Process No.~2019/00763-0 and No.~2023/07313-6.
This work was in the projects of Instituto Nacional de Ci\^{e}ncia e
Tecnologia - Nuclear Physics and Applications
(INCT-FNA), Brazil, Process No.~464898/2014-5.
\vspace{-2ex}

%

\begin{thebibliography}{100}
%
%
\bibitem{Zeminiani:2023gqc}
G.~N.~Zeminiani, S.~L.~P.~G.~Beres and K.~Tsushima,
Phys. Rev. D \textbf{110}, 094045 (2024).

\bibitem{Tsushima:1998ru}
K.~Tsushima, D.~H.~Lu, A.~W.~Thomas, K.~Saito and R.~H.~Landau,
Phys. Rev. C \textbf{59}, 2824 (1999).

\bibitem{Krein:2017usp}
G.~Krein, A.~W.~Thomas and K.~Tsushima,
Prog. Part. Nucl. Phys. \textbf{100}, 161 (2018).


\bibitem{Krein:2010vp}
G.~Krein, A.~W.~Thomas and K.~Tsushima,
Phys. Lett. B \textbf{697}, 136 (2011).

\bibitem{Cobos-Martinez:2020ynh}
J.~J.~Cobos-Mart\'\i{}nez, K.~Tsushima, G.~Krein and A.~W.~Thomas,
Phys. Lett. B \textbf{811}, 135882 (2020).

\bibitem{Zeminiani:2024dyo}
G.~N.~Zeminiani, J.~J.~Cobos-Mart{\'\i}nez and K.~Tsushima,
Phys. Rev. C \textbf{111}, 055202 (2025).


\bibitem{Cobos-Martinez:2025iqg}
J.~J.~Cobos-Mart{\'\i}nez, G.~N.~Zeminiani and K.~Tsushima,
Symmetry \textbf{17}, 787 (2025).

\bibitem{Zeminiani:2020aho}
G.~N.~Zeminiani, J.~J.~Cobos-Martinez and K.~Tsushima,
Eur. Phys. J. A \textbf{57}, 259 (2021).







\bibitem{Guichon:1987jp}
P.~A.~Guichon,
Phys.\ Lett.\ B \textbf{200}, 235 (1988).


\bibitem{Saito:2005rv}
K.~Saito, K.~Tsushima and A.~W.~Thomas,
Prog. Part. Nucl. Phys. \textbf{58}, 1 (2007).


\bibitem{Tsushima:1997df}
K.~Tsushima, K.~Saito, A.~W.~Thomas and S.~V.~Wright,
Phys. Lett. B \textbf{429}, 239 (1998)
[erratum: Phys. Lett. B \textbf{436}, 453 (1998)]

\bibitem{Guichon:1995ue}
P.~A.~M.~Guichon, K.~Saito, E.~N.~Rodionov and A.~W.~Thomas,
Nucl. Phys. A \textbf{601}, 349 (1996).


\bibitem{Tsushima:2020gun}
K.~Tsushima,
PTEP \textbf{2022}, 043D02 (2022).


\bibitem{ParticleDataGroup:2022pth}
R.~L.~Workman \textit{et al.} [Particle Data Group],
PTEP \textbf{2022}, 083C01 (2022).


\bibitem{Martin-Gonzalez:2022qwd}
B.~Mart\'\i{}n-Gonz\'alez, P.~G.~Ortega, D.~R.~Entem, F.~Fern\'andez and J.~Segovia,
Phys. Rev. D \textbf{106}, 054009 (2022).


\bibitem{Lin:2000ke}
Z.~w.~Lin and C.~M.~Ko,
Phys. Lett. B \textbf{503}, 104 (2001).




\bibitem{Lodhi:2007zz}
M.~A.~K.~Lodhi and R.~Marshall,
Nucl. Phys. A \textbf{790}, 323 (2007).



%





\end{thebibliography}
%
%


\vspace{-2ex}

\end{document}